\documentclass[conference]{IEEEtran}
\usepackage{cite}
\usepackage{amsmath,amssymb,amsfonts}
\usepackage[ruled,vlined]{algorithm2e}
\usepackage[noend]{algcompatible}
\usepackage{graphicx}
\usepackage{textcomp}
\usepackage{xcolor}
\usepackage{fancyhdr}
\usepackage[hyphens]{url}
\usepackage{tikz}
\usepackage{soul}
\usepackage{listings}
\newcommand*\circled[1]{\tikz[baseline=(char.base)]{
            \node[shape=circle,fill,inner sep=1pt] (char) {\textcolor{white}{#1}};}}

\def\BibTeX{{\rm B\kern-.05em{\sc i\kern-.025em b}\kern-.08em
    T\kern-.1667em\lower.7ex\hbox{E}\kern-.125emX}}

\fancypagestyle{firstpage}{
  \fancyhf{}

  \fancyhead[C]{\normalsize{Preprint. Under review.}} 
  \fancyfoot[C]{\thepage}
}

\newcommand*{\affaddr}[1]{#1} 
\newcommand*{\affmark}[1][*]{\textsuperscript{#1}}
\newcommand*{\email}[1]{\texttt{#1}}

\title{
MPU: Towards Bandwidth-abundant SIMT Processor via Near-bank Computing
}
\author{%
Xinfeng Xie$^{*}$\affmark[1], Peng Gu$^{*}$\affmark[1], Yufei Ding\affmark[2], Dimin Niu\affmark[3], Hongzhong Zheng\affmark[3], Yuan Xie\affmark[1,3]\\
\affaddr{\affmark[1]Department of Electrical and Computer Engineering, UCSB, Santa Barbara, USA}\\
\affaddr{\affmark[2]Department of Computer Science, UCSB, Santa Barbara, USA}\\
\email{Email:\{xinfeng,peng\_gu,yufeiding,yuanxie\}@ucsb.edu}\\
\affaddr{\affmark[3]Alibaba DAMO Academy, Sunnyvale, USA}\\
\email{Email:\{dimin.niu,hongzhong.zheng\}@alibaba-inc.com}
\thanks{$^{*}$Xinfeng Xie and Peng Gu are co-primary authors.}
}
\IEEEoverridecommandlockouts
\begin{document}
\maketitle
\thispagestyle{firstpage}
\pagestyle{plain}


\begin{abstract}

With the growing number of data-intensive workloads, GPU, which is the state-of-the-art single-instruction-multiple-thread (SIMT) processor, is hindered by the memory bandwidth wall.
To alleviate this bottleneck, previously proposed 3D-stacking near-bank computing accelerators benefit from abundant bank-internal bandwidth by bringing computations closer to the DRAM banks.
However, these accelerators are specialized for certain application domains with simple architecture data paths and customized software mapping schemes. 
For general purpose scenarios, lightweight hardware designs for diverse data paths, architectural supports for the SIMT programming model, and end-to-end software optimizations remain challenging.

To address these issues, we propose MPU (\underline{M}emory-centric \underline{P}rocessing \underline{U}nit), the first SIMT processor based on 3D-stacking near-bank computing architecture.
First, to realize diverse data paths with small overheads while leveraging bank-level bandwidth, MPU adopts a hybrid pipeline with the capability of offloading instructions to near-bank compute-logic.
Second, we explore two architectural supports for the SIMT programming model, including a near-bank shared memory design and a multiple activated row-buffers enhancement.
Third, we present an end-to-end compilation flow for MPU to support CUDA programs.
To fully utilize MPU's hybrid pipeline, we develop a backend optimization for the instruction offloading decision.
The evaluation results of MPU demonstrate $3.46\times$ speedup and $2.57\times$ energy reduction compared with an NVIDIA Tesla V100 GPU on a set of representative data-intensive workloads.

\end{abstract}

\setlength{\textfloatsep}{10pt}
\section{Introduction}\label{sec:intro}
Nowadays, parallel computing platforms play an increasingly important role in accelerating emerging data-intensive applications from various domains such as deep learning~\cite{lym2019prunetrain}, image processing~\cite{gu2020ipim}, and bioinformatics~\cite{huangfu2019medal}.
In particular, a general purpose graphics processing unit (GPGPU) with its single-instruction-multiple-thread (SIMT) programming model benefits these workloads by exploiting massive memory-level parallelism and providing DRAM bandwidth higher than traditional CPUs.
Despite the success of GPGPU in accelerating these workloads, its further performance scaling is constrained by the "memory bandwidth wall"~\cite{rogers2009scaling} challenge.
From the application's perspective, workloads from these data-intensive applications usually require a large number of memory accesses with little computation.
This characteristic makes memory bandwidth the dominating factor of performance.
Unfortunately, from the technology's point of view, the scaling of main memory bandwidth for the compute-centric architecture is confined by both the insufficient number of off-chip I/O pins~\cite{liu2010understanding} and the expensive data movement energy~\cite{horowitz20141}.
To further corroborate this bandwidth-bound behavior, in Sec.\ref{sec:motivation} we evaluate a set of representative benchmarks on an NVIDIA Tesla V100 GPU~\cite{Ref:GPU}, and find that on average the off-chip memory bandwidth utilization ($55.90\%$) overwhelmingly surpasses the ALU utilization ($2.57\%$) as shown in Fig.~\ref{fig:motivation}.

The 3D-stacking near-data-processing (3D-NDP) architecture~\cite{balasubramonian2014near} emerges as a promising approach to alleviate this memory bandwidth bottleneck.
Currently, high-end GPUs~\cite{macri2015amd,o2014highlights} are equipped with high-bandwidth memory (HBM) stacks~\cite{sohn20171}, where off-chip data transfers need to go through the low performance I/Os on the silicon interposer.
The principal idea of 3D-NDP is to closely integrate affordable logic components adjacent to the memory stack.
A large number of pioneering studies have adopted the processing-on-base-logic-die (PonB) architecture, where general purpose cores (e.g., SIMT cores~\cite{zhang2014top,hsieh2016transparent,pattnaik2016scheduling,kersey2017lightweight,kim2017toward,wen2017optimizing,nai2018coolpim,wu2020tuning}) are placed on the base logic die of the 3D stack to benefit from the intra-stack bandwidth enhancement (around $2\times$ w.r.t. HBM~\cite{wu2020tuning}). 
This solution provides a mediocre bandwidth improvement because intra-stack memory accesses are still bounded by the limited number of through-silicon-vias (TSVs) between memory dies and the base logic die.
To overcome this bandwidth bottleneck of TSVs, recent near-bank accelerators~\cite{yazdanbakhsh2018dram,shin2018mcdram,aga2019co,gu2020ipim} further move simple arithmetic units closer to the DRAM banks to harvest the abundant bank-internal bandwidth (around $10\times$ w.r.t. process-on-logic-die solution~\cite{gu2020ipim}).
These near-bank accelerators have demonstrated significant speedups (around $2\times-14\times$ w.r.t. GPU) thus they are promising to tackle the GPU's memory bandwidth issue.

Despite its potential to provide plentiful memory bandwidth, there are still several challenges for near-bank computing in accelerating general purpose data-intensive workloads.
First, the pipeline of SIMT processors contains complex logic components (e.g., load-store-unit~\cite{nyland2011systems}) and large register files~\cite{lindholm2008simulating}.
Different from the prior near-bank accelerators customized for applications, the SIMT pipeline is needed for general purpose data-intensive workloads.
Naively placing the whole pipeline with complex logic components and complicated data paths in the DRAM die introduces an intolerable area overhead~\cite{kogge1994execube,patterson1997case,draper2002architecture}.
Second, the efficient support of the SIMT programming model on near-bank computing is needed, especially the inter-thread communication and the dynamic scheduling of warps~\cite{kirk2007nvidia}.
As the shared memory is frequently used for inter-thread communication in a thread block, directly placing it on the base logic die will incur enormous TSV traffic.
The dynamic scheduling of warps could disrupt the row-buffer locality of DRAM banks, seriously downgrading bandwidth utilization.
Third, it requires both the end-to-end support of a parallel programming language to ease the programmers' burden and backend compiler optimizations for near-bank computing to exploit hardware potentials.

To address these challenges, we design MPU (Memory-centric Processing Unit), the first SIMT processor based on 3D-stacking near-bank computing architecture, and an end-to-end compiler flow supporting CUDA programs~\cite{nvidia2007compute} with optimizations tailored to MPU.
First, we design a hybrid SIMT pipeline for MPU where only a small number of registers and other lightweight components are added on the DRAM dies.
At runtime, instructions are fetched, decoded, and issued on the base logic die while they can be offloaded to near-bank units (NBU) according to either compiler hints or hardware policies.
To facilitate this hybrid execution of instructions, we propose an instruction offload engine to make instruction movement decisions, a register track table and a register move engine to flexibly transfer registers, and a load-store unit extension to handle near-bank load/store requests.
Second, we propose two architectural optimizations for the SIMT model.
For the shared memory, we move it to the DRAM die and restructure the core organization by placing all NBUs associated with the same core on the same DRAM die.
For the dynamic scheduling of thread warps, we enable multiple activated row-buffers per DRAM bank to reduce the ping-pong effect thus improving the bandwidth.
Third, we propose an end-to-end compilation flow to support CUDA programs.
To optimize instruction offloading location on MPU's hybrid pipeline, we further propose a novel instruction and register location annotation algorithm through the static analysis of instructions, which effectively reduces the data movement among the shared TSVs.

The contributions of our work are summarized as follows:
\begin{itemize}
	\item To the best of our knowledge, we design the first near-bank SIMT processor using a hybrid pipeline with an instruction offloading mechanism. By integrating lightweight hardware components on the DRAM die, MPU achieves a small area overhead for general purpose processing.
	\item We propose two architectural optimizations for the SIMT model, including the near-bank shared memory to reduce data movement and multiple activated row-buffers to alleviate ping-pong effects in the dynamic warp scheduling.
	\item We develop an end-to-end compilation flow supporting CUDA programs on MPU and a novel backend optimization annotating the locations of registers and instructions.
	\item Evaluation results of representative data-intensive workloads show that MPU with all optimizations achieves $3.46\times$ speedup and $2.57\times$ energy reduction on average over an NVIDIA Tesla V100 GPU. 
\end{itemize}

\section{Motivation}\label{sec:motivation}

\begin{figure}[!t]
  \centering
  \includegraphics[width=1\linewidth]{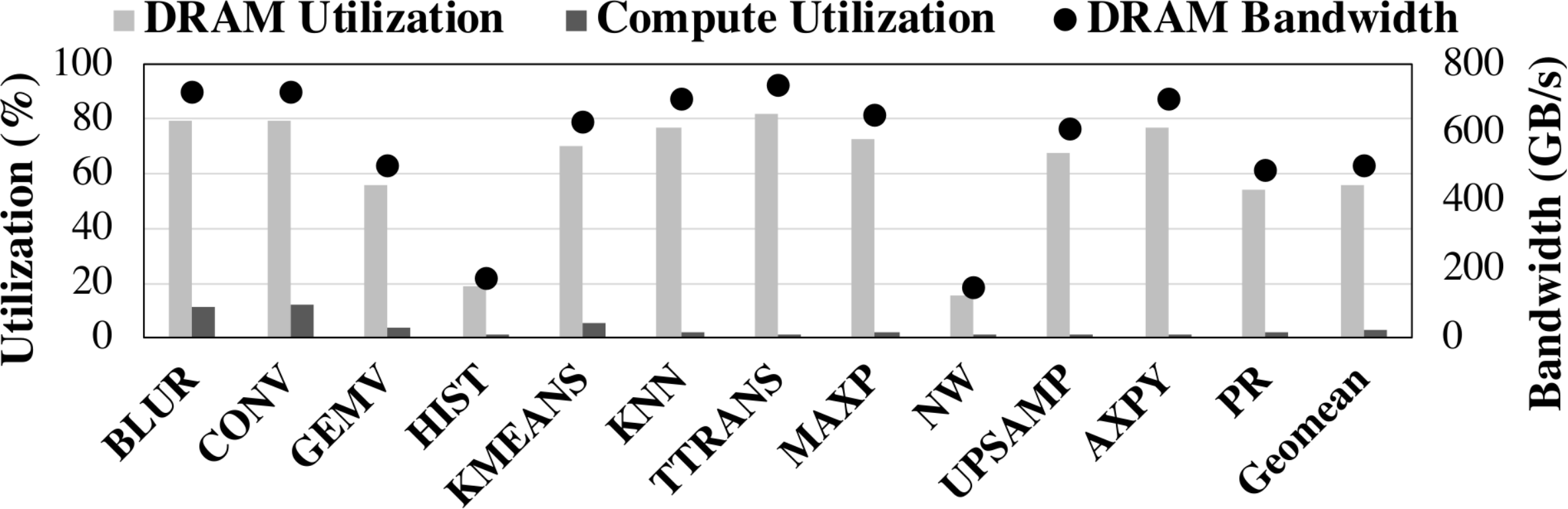}
  \caption{Profiling results for data-intensive workloads on NVIDIA Tesla V100 GPU.}
  \label{fig:motivation}
\end{figure}

\begin{figure*}[!h]
  \centering
  \includegraphics[width=1\linewidth]{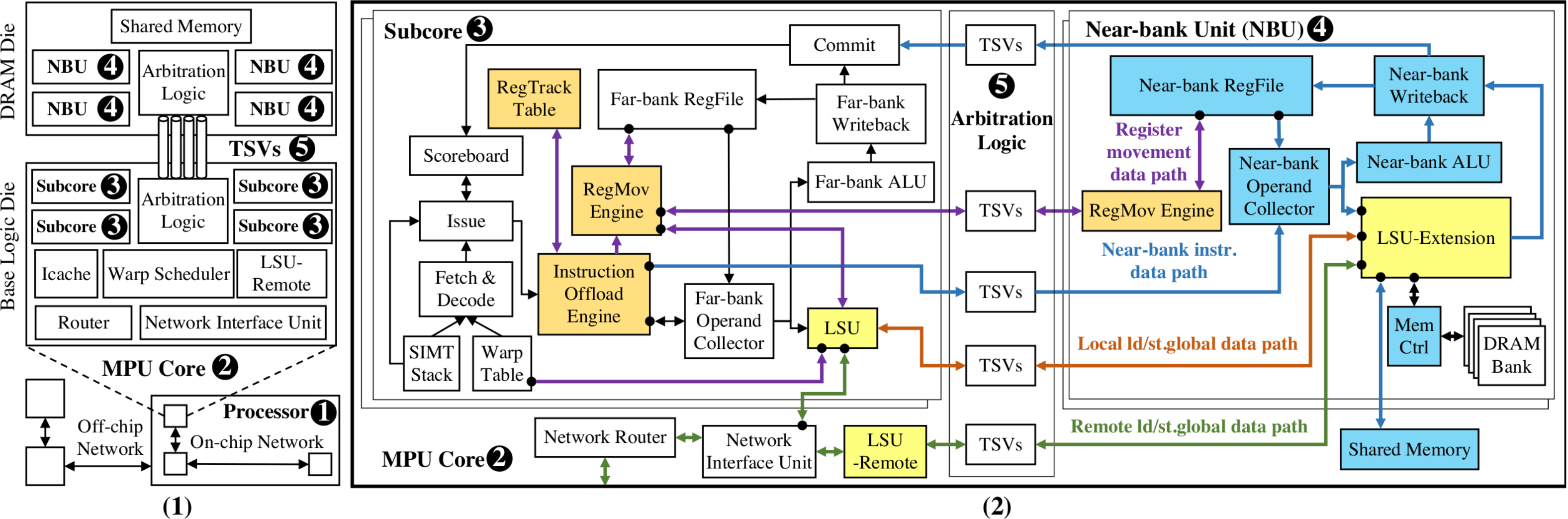}
  \caption{(1) MPU architecture overview. (2) Detailed microarchitecture of a MPU core (left: Subcores, middle: TSVs, right: NBUs). Added near-bank components are colored in blue. Newly supported components for instruction offloading and register movement (Sec.\ref{subsec:arch:instr_offld}) are colored in orange. Hybrid load-store unit (LSU) components (Sec.\ref{subsec:arch:hybrid_lsu}) are colored in yellow.}
  \vspace{-1em}
  \label{fig:arch}
\end{figure*}

Despite the success of the graphics processing unit (GPU) in accelerating data-intensive parallel programs, we observe from performance characterizations that the memory bandwidth is a serious performance challenge for these workloads on the state-of-the-art GPU.
Specifically, we evaluate a set of representative data-intensive workloads from various application domains including deep learning, bioinformatics, linear algebra, and image processing applications as detailed in Table~\ref{tab:exp:benchmark}.
The measured memory bandwidth, bandwidth utilization, and compute utilization of an NVIDIA Tesla V100 GPU~\cite{martineau2018benchmarking} are shown in Fig.\ref{fig:motivation}.
On average, these benchmarks achieve $55.90\%$ DRAM bandwidth utilization ($503.10$ GB/s) and $2.57\%$ compute utilization.
The saturation of DRAM bandwidth and the low utilization of the compute resources exhibit a memory-bandwidth bound behavior.
This performance characteristic results from the low arithmetic density and the regular memory access patterns in most of these workloads. 
Also, we note that the workloads $HIST$ and $NW$ show relatively low bandwidth utilization as a result of the long memory access latency on GPU.

For workloads suffering from either the limited DRAM bandwidth or the long DRAM access latency on GPU, near-bank computing is a promising architecture to alleviate these performance bottlenecks because of both abundant bank-level memory bandwidth and reduced memory access latency.
However, prior near-bank computing accelerators~\cite{yazdanbakhsh2018dram,shin2018mcdram,aga2019co,gu2020ipim} are domain-customized, since they have simple data paths, application-specific mapping strategies, and inefficient general purpose programming language support.
The lack of programmability for these accelerators confines them to a niche application market, adding non-recurring engineering costs in manufacturing.
Moreover, parallel data-intensive workloads usually come from various application domains, making none of these near-bank accelerators feasible to support all of these parallel programs.

In summary, the memory bandwidth bottleneck on the state-of-the-art GPU urges the need for a higher memory bandwidth for data-intensive parallel programs, and the huge overheads of placing an SIMT processor near banks introduce unique technical challenges.
Both of these factors motivate us to design MPU, the first general purpose SIMT processor based on 3D-stacking near-bank computing to exploit bank-level bandwidth and alleviate the GPU bandwidth bottleneck.
\section{Background}~\label{sec:bg}~\label{subsec:bg:nearbank}~\label{subsec:bg:simt}

\noindent \textbf{3D-stacking Near-data Computing (3D-NDC):}
3D-NDC is based on commercially available 3D-stacking memory modules (HBM~\cite{sohn20171} and HMC~\cite{leidel2016hmc}) with DRAM dies stacked on the top of a base logic die.
Previous GPU-style processing-on-logic-die solutions~\cite{zhang2014top,hsieh2016transparent,pattnaik2016scheduling,kersey2017lightweight,kim2017toward,wen2017optimizing,nai2018coolpim,wu2020tuning} usually place simple SIMT cores (Sec.\ref{subsec:bg:simt}) on the base logic die, and the cores access memory using TSVs (Through-Silicon-Vias)~\cite{xie2015stacking}, which are vertical interconnects shared among 3D layers.
However, this solution is limited by the available bandwidth provided by TSVs (currently $307GB/s$ for one stack with 1024 TSVs~\cite{sohn20171}), and scaling TSVs is very difficult due to the large area overhead (already $18.8\%$ of each 3D layer~\cite{sohn20171}).
To solve the TSV challenge, recent work pushes simple compute logic adjacent to each bank to utilize the enormous bank-level bandwidth for domain-specific acceleration~\cite{yazdanbakhsh2018dram,shin2018mcdram,aga2019co,gu2020ipim}.
However, it is challenging to support general purpose near-bank architecture due to the large area overhead of fabricating general purpose cores in DRAM dies.
We solve this challenge by a novel hybrid architecture that decouples the SIMT pipeline and duplicates some parts of the backend stages onto the near-bank DRAM dies (Sec.\ref{subsec:arch:hybrid_pipeline}). 

\noindent \textbf{Single-instruction-multiple-thread (SIMT) Processor:}
SIMT model~\cite{nemirovsky2013multithreading} is widely adopted in modern GPUs~\cite{macri2015amd,o2014highlights} for accelerating parallel computing programs.
The pipeline can be divided into front-pipeline which consists of instruction fetch, decode, and issue stages, middle-pipeline which consists of execution and writeback stages, and end-pipeline which consists of the commit stage.
The front-pipeline and the end-pipeline usually involve complex control and logic circuits, such as warp table, SIMT stack, scoreboard, and load-store unit.
The middle-pipeline contains arithmetic instruction and memory access instructions.
MPU reduces the near-bank overhead by duplicate some part of the middle-pipeline to DRAM dies, and add corresponding instruction offloading mechanisms as detailed in Sec.\ref{subsec:arch:instr_offld}.
The SIMT model also has some special features that require optimizations.
First, the shared memory is extensively used for inter-thread communication inside the same thread block.
Placing it in the base logic die may introduce extra communication traffic among the 3D layers.
Second, the SIMT scheduling causes warps to access different row-buffers interchangeably, causing a row-buffer ping-pong effect.
MPU explores architectural optimizations, near-bank shared memory and multiple activated row-buffers, for these two features detailed in Sec.\ref{subsec:arch:simt_opt}.

\section{Architecture Design}~\label{sec:arch}
First, we discuss MPU's high-level design in Sec.\ref{subsec:arch:microarch}. 
Second, we introduce MPU's hybrid pipeline in Sec.\ref{subsec:arch:hybrid_pipeline}, and further describe its instruction offloading mechanism in Sec.\ref{subsec:arch:instr_offld} and hybrid load-store unit in Sec.\ref{subsec:arch:hybrid_lsu}.
Next, we present two architectural optimizations for SIMT model in Sec.\ref{subsec:arch:simt_opt}.

\subsection{Microarchitecture Overview}~\label{subsec:arch:microarch}
From the high-level, MPU adopts a scalable design with many processors (Fig.\ref{fig:arch}\circled{1}) interconnected by an off-chip network (similar to SERDES links in the HMC~\cite{leidel2016hmc}) as shown in the bottom of Fig.\ref{fig:arch} (1).
Each processor is a 3D-stacking cube of a base logic die stacked with multiple DRAM dies, connected by vertically shared buses called the through silicon vias (TSVs)~\cite{xie2015stacking} (Fig.\ref{fig:arch}\circled{5}).
The base logic die is horizontally partitioned into an array of SIMT cores (Fig.\ref{fig:arch}\circled{2}) interconnected by an on-chip 2D-mesh network~\cite{jiang2013detailed}.

To harvest the near-bank bandwidth with small overheads on DRAM dies, MPU's SIMT core adopts a hybrid pipeline design.
In the MPU core (Fig.\ref{fig:arch}\circled{2}), complex logics are placed on the base logic die, and some lightweight components in the execution stage are replicated on near-bank locations.
On the base logic die, a core consists of four subcores (Fig.\ref{fig:arch}\circled{3}), an instruction cache, a warp scheduler, and components for handling inter-core traffic (network interface unit, router, and LSU-Remote).
The TSVs (Fig.\ref{fig:arch}\circled{5}) are evenly divided among the cores ($64b$ data buses per core), via which the subcores can communicate with near-bank components on DRAM dies.
All the core's near-bank components are located within the same DRAM die, containing four near-bank units (NBUs) (Fig.\ref{fig:arch}\circled{4}) and the shared memory.
To enable efficient processing in this hybrid architecture (Sec.\ref{subsec:arch:hybrid_pipeline}), we propose a novel instruction offloading mechanism and a hybrid load-store unit design.

In addition, MPU considers two architectural optimizations for the SIMT programming model in Sec.\ref{subsec:arch:simt_opt}.
First, we find that naively implementing shared memory on the base logic die results in poor performance, so we restructure the core's 3D organization and develop a near-bank shared memory design.
Second, we observe that the dynamic execution of SIMT warps may disrupt the row-buffer locality of DRAM banks, so we adopt a technique to enable multiple activated row-buffers inside the same DRAM bank.

\subsection{Hybrid Pipeline}~\label{subsec:arch:hybrid_pipeline}
As illustrated in Fig.\ref{fig:arch} (2), an MPU core (Fig.\ref{fig:arch} (2)\circled{2}) adopts a hybrid pipeline design that is split between the base logic die (subcore) (Fig.\ref{fig:arch} (2)\circled{3}) and the DRAM die (near-bank unit, NBU) (Fig.\ref{fig:arch} (2)\circled{4}).
The frontend components of the SIMT pipeline mostly comprise of control flow and data dependency logic, so they are mainly contained in the subcores, including instruction fetch (I-cache, SIMT stack, warp table), decode, and issue (scoreboard) stages.
For the backend pipeline, MPU duplicates some simple parts from the subcore to the NBU, including the register file, operand collectors, and ALUs.
Other complex units such as LSU and network interface units are left on the base logic die.
Note that the memory controller and the shared memory are entirely moved from the base logic die to the DRAM die, since near-bank memory controller will reduce TSV traffic for DRAM commands~\cite{yazdanbakhsh2018dram,gu2020ipim}, and shared memory can reduce TSV traffic for register movement (Sec.\ref{subsec:arch:simt_opt}).

In addition to the original pipeline components, to assist flexible instruction offloading, each subcore also adds an instruction offload engine, a register move engine, and an associated register track table (Sec.\ref{subsec:arch:instr_offld}).
Besides, the load-store unit (LSU) is modified and augmented to support remote data traffic and near-bank instruction offloading (sec.\ref{subsec:arch:hybrid_lsu}).

\begin{figure}[!t]
  \centering
  \includegraphics[width=1\linewidth]{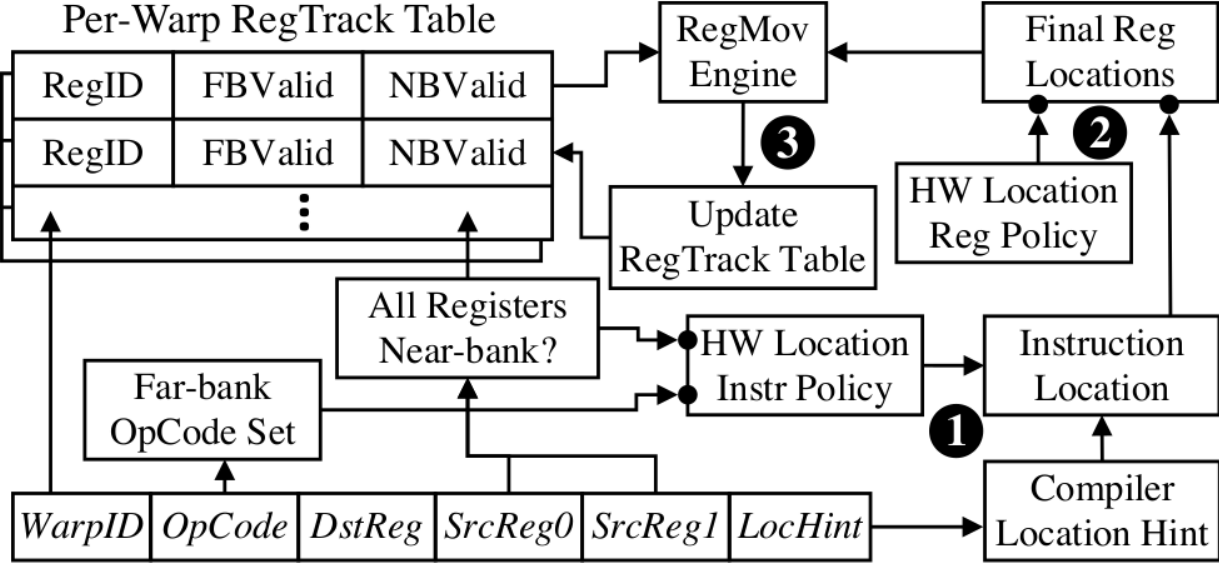}
  \caption{Instruction Offloading Mechanism.}
  \label{fig:instr_offld}
\end{figure}

\subsubsection{Instruction Offloading Mechanism}~\label{subsec:arch:instr_offld}

The instruction offload engine after the issue stage will decide whether to offload the instruction for near-bank execution (\textit{Near-bank instr. data path} in Fig.\ref{fig:arch} (2)).
The offloaded instruction will first travel through the TSVs, then access the near-bank operand collector to collect operand data from the near-bank register file.
Then, the arithmetic and logic computation instructions will be sent to the near-bank ALUs, and the $ld/st$ instructions will be provided to the LSU-Extension for further processing, where the shared memory or the DRAM controller is involved.
After the execution finishes, the resulting registers will be written back into the near-bank register file, and the instruction is returned to the subcore for the final commit, where the scoreboard clears its data dependency.

The first step (Fig.\ref{fig:instr_offld}\circled{1}) is to identify the target instruction's location according to three policies with decreasing priority.
The first policy will set the instruction with the far-bank location if the corresponding operation type ($OpCode$) falls in the far-bank operation set.
For example, since address range checking and memory coalescing can only be performed by the LSU in the subcore, currently $ld/st.global$ instructions are classified as far-bank locations.
If the first hardware policy cannot determine the location, then the compiler hint associated with the instruction will determine whether this instruction will be offloaded to NBU or not.
If the instruction has no compiler hints, then the hardware will check the register track table. 
The instruction will be offloaded to NBU if all source registers have valid near-bank copies.
Otherwise, the instruction will be passed to the far-bank.
This default policy takes the far-bank subcore as a fall-back location for all instructions as it has the full pipeline support.

The second step (Fig.\ref{fig:instr_offld}\circled{2}) is to determine the locations for the source registers using the hardware policy or the instruction location derived in the first step.
For $ld/st.global$, the hardware policy will set the address register location to be far-bank, since it is required by the LSU. The data register location is set to near-bank.
For $ld/st.shared$, both the address register and the data register are set to near-bank location. 
If the hardware policy is not given, all the source and destination registers' locations will follow the instruction's location.

The third step (Fig.\ref{fig:instr_offld}\circled{3}) will move registers to their corresponding locations if they are not currently available in the register track table (\textit{Register Movement data path} in Fig.\ref{fig:arch} (2)).
For example, the instruction offload engine may require register $\%r1$ to be valid in the far-bank register file, while the register track table indicates $\%r1$ only exists in the near-bank register file (e.g., \textit{FBValid} is False but \textit{NBValid} is True).
Then, the far-bank register move engine initiates a request to the near-bank register move engine, which then reads $\%r1$ from the near-bank register file and returns it to the far-bank register move engine.
The far-bank register move engine will write $\%r1$ into the far-bank register file, and the instruction offloading engine will start instruction offloading once all registers are in the target locations.
Note that we optimize register locations in Sec.\ref{sec:compiler}, so a hit in the register track table will not cause register movement.
In the end, the register track table is updated to reflect the most current register location information.

\subsubsection{Hybrid Load-store Unit (LSU)}~\label{subsec:arch:hybrid_lsu}
\begin{figure}[!t]
  \centering
  \includegraphics[width=1\linewidth]{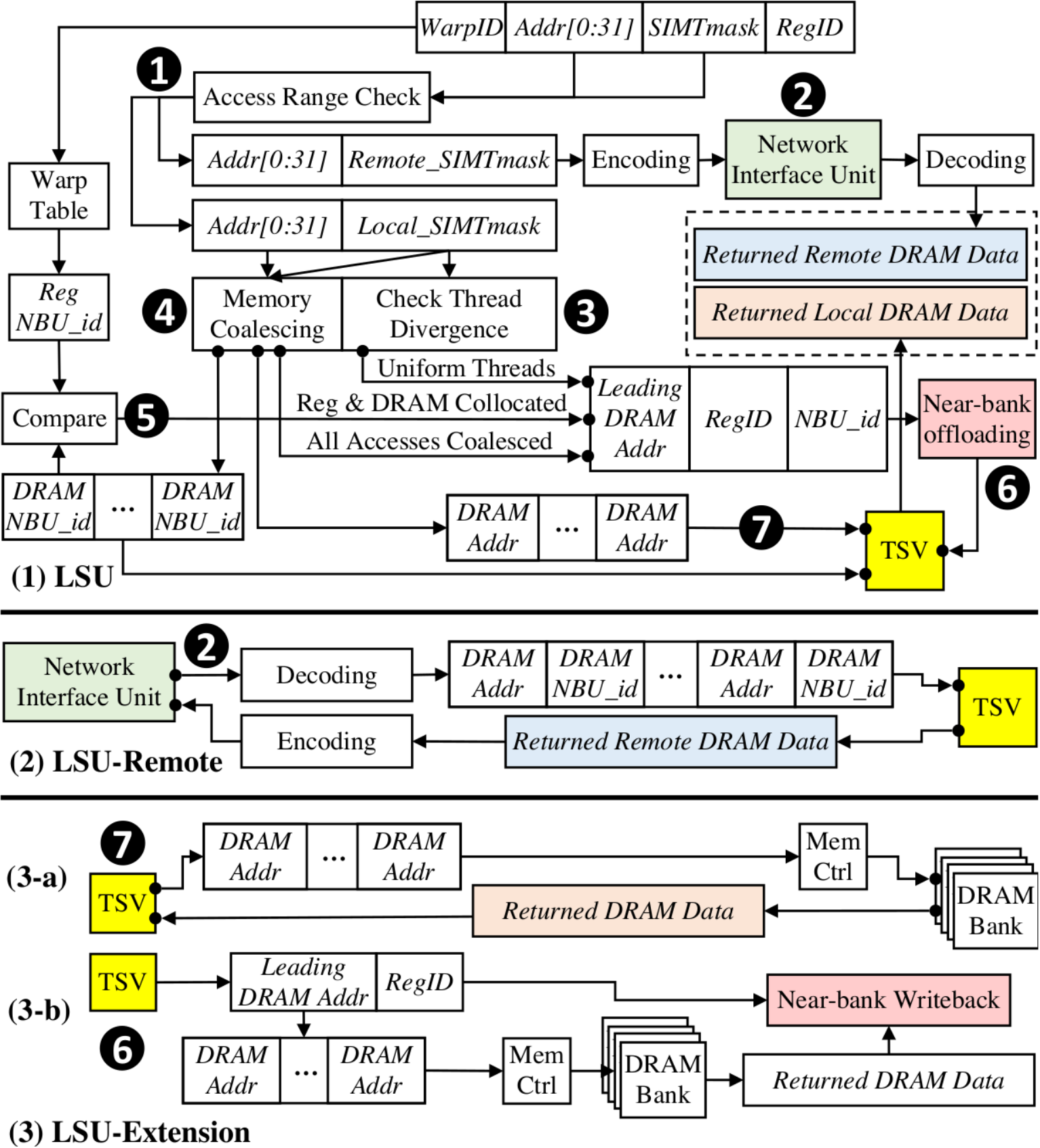}
  \caption{Three load-store unit (LSU) components' microarchitecture using $ld.global$ data path as an example. (1) LSU in each subcore, (2) LSU-Remote in each core, (3) LSU-Extension in each near-bank unit}
  \label{fig:hybrid_lsu}
\end{figure}

The original LSU in each subcore is augmented with the LSU-Remote in each core to handle remote traffic (\textit{Remote $ld/st.global$ data path} in Fig.\ref{fig:arch} (2)), and the LSU-Extension in each NBU to handle both near-bank instruction offloading and local DRAM transactions (\textit{Local $ld/st.global$ data path} in Fig.\ref{fig:arch} (2)).
We will introduce them and use $ld.global$ instruction as an example in the following paragraphs.

\textbf{LSU:}
As shown in Fig.\ref{fig:hybrid_lsu} (1), after receiving a $ld.global$ instruction, the LSU first performs address range checking (Fig.\ref{fig:hybrid_lsu}\circled{1}) to split the instruction to remote access and local access.
If there is remote access, it is encoded and sent to the network interface unit (Fig.\ref{fig:hybrid_lsu}\circled{2}) to request remote data.
For the local access, the following steps are performed concurrently.
First, the LSU will check if all the SIMT mask fields are valid or not to determine thread divergence (Fig.\ref{fig:hybrid_lsu}\circled{3}).
Second, the LSU will perform memory coalescing (Fig.\ref{fig:hybrid_lsu}\circled{4}) on the local memory addresses.
It will also judge if all the memory addresses are perfectly coalesced, meaning that the load request will access a continuous DRAM address space.
Third, the LSU will compare the $NBU\_id$ field of the generated DRAM addresses with the $NBU\_id$ associated with the register in the given warp (Fig.\ref{fig:hybrid_lsu}\circled{5}).
The $ld.global$ instruction is decided for near-bank offloading (Fig.\ref{fig:hybrid_lsu}\circled{6}) only if all threads are valid, register and DRAM addresses have the same $NBU\_id$, and all DRAM addresses are coalesced.
Note that since all of the above assumptions are satisfied, we only need to transfer the leading DRAM address, register ID, and the $NBU\_id$. 
The SIMB mask will be ignored and restored in the LSU-Extension side.
If the above conditions cannot be met, the LSU will issue DRAM transactions to the LSU-Extension (Fig.\ref{fig:hybrid_lsu}\circled{7}) and gather returned local DRAM data.
Combined with the returned remote DRAM data, the final DRAM data will be used to compose a register write request and transferred for near-bank writeback.
The reason to load the DRAM data first to the near-bank register file is that it can benefit near-bank execution due to the reduction of TSV traffic.
For far-bank execution, the register data will eventually be brought down to the far-bank register file, so there is no increase in the TSV traffic.

\textbf{LSU-Remote:}
As shown in Fig.\ref{fig:hybrid_lsu} (2), LSU-Remote receives remote $ld.global$ request and decodes them into a series of DRAM addresses and DRAM $NBU\_id$.
It then sends these DRAM transactions to the LSU-Extension through the TSVs.
After receiving the returned DRAM data, it encodes it together with the source core location and request ID and composes a response packet, finally sending it back to the original core's LSU who requests this DRAM data.

\textbf{LSU-Extension:}
As shown in Fig.\ref{fig:hybrid_lsu} (3), LSU-Extension has two data paths.
In Fig.\ref{fig:hybrid_lsu} (3-a), it handles DRAM transaction requests from the TSVs by sending the DRAM addresses to the memory controller, and sends back the returned DRAM data through TSV either to the LSU-Remote or the LSU.
In Fig.\ref{fig:hybrid_lsu} (3-b), it handles offloaded local $ld.global$ instruction by first restoring the full address list from the leading DRAM address.
Then, it sends the DRAM addresses to the memory controller, gathers returned DRAM data, and stores the data into the near-bank register file according to the register ID.

\subsection{Optimizations for the SIMT model}~\label{subsec:arch:simt_opt}

\begin{figure}[!t]
  \centering
  \includegraphics[width=1\linewidth]{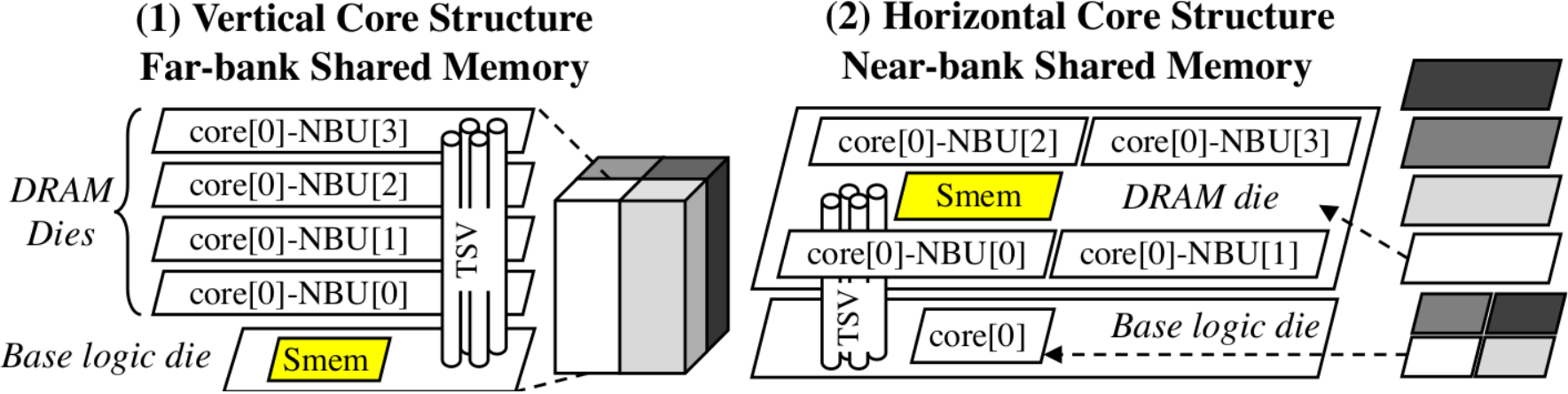}
  \caption{(1) Vertical and (2) horizontal core structure.}
  \label{fig:smem_arch}
\end{figure}

\textbf{Near-bank Shared Memory Design:}
The shared memory is extensively used for inter-thread communication in the same thread block for a great number of important GPU benchmarks~\cite{kirk2007nvidia}.
If the default shared memory location is set in the far-bank subcore on the base logic die, a lot of register data movement traffic will be created and the TSVs will be congested, causing significant performance loss.
Thus, it will be desirable to enable a near-bank shared memory design and set the default register location for $ld/st.shared$ to the near-bank register file.
However, this is impossible in the vertical core structure (Fig.\ref{fig:smem_arch} (1)) in the default HMC-style setting~\cite{leidel2016hmc}, where all the NBUs associated with a core are distributed among multiple 3D stacks.
Under such an assumption, moving the shared memory ($Smem$) to each NBUs means that the shared memory is split into each 3D layer and inter-thread shared memory accesses need to go through the bandwidth-bound TSVs.
To enable the near-bank shared memory, we restructure the core's 3D-organization as shown in the horizontal core design in Fig.\ref{fig:smem_arch} (2).
In our solution, we put all NBUs of the same core into the same DRAM die, so that all NBUs can access the near-bank shared memory without TSV's constraints.
In Sec.\ref{subsec:exp:hw_analysis}, we confirm the benefits of this optimization on benchmarks that extensively use shared memory.

\textbf{Multiple Activated Row-Buffers Design:}
The dynamic execution of warps will create a ping-pong effect on DRAM's row-buffer.
Ideally, warps from the same thread block executing the same memory access instruction will have continuous DRAM addresses and result in a high row-buffer hit rate.
However, the hardware dynamically issues available instructions from each warp, resulting in the ping-pong effect of different warps accessing a few row-buffers irregularly.
Since MPU has no hardware cache, this ping-pong effect will cause frequent DRAM precharge and activations, significantly downgrade its performance.

To solve this issue in a software transparent way, we observe that for a lot of benchmarks we evaluate, only a small set of row-buffers are active in a short period.
If we can enable multiple row-buffers to be simultaneously activated, then this ping-pong effect will be greatly alleviated.
Based on the design of MASA (Multitude of Activated Subarrays)~\cite{kim2012case} which enables multiple subarrays' row-buffers to be activated in parallel, we change our address mapping so that continuous DRAM row addresses will be mapped to interleaved subarrays' physical row.
Extra row address latches and access transistors are added to enable different warps to access different row buffers in independent subarrays.
Through evaluations in Sec.\ref{subsec:exp:hw_analysis}, we confirm that this design can decrease the row-buffer miss rate and increase performance for multiple benchmarks.
\section{Programming Interface and Compiler}\label{sec:compiler}

\begin{lstlisting}[float=tp, language=C++, caption={Code example of scalar-vector multiplication.}, emph={Var, Func}, emphstyle={\color{blue}\ttfamily}, label={lst:example}, basicstyle=\footnotesize]
// CUDA kernel for scalar-vector multiplication
__global__ void ScalarVectorMultiply(float* input, 
    float* output, float alpha, int len) {
    int numThreads = gridDim.x * blockDim.x;
    int tid = blockIdx.x * blockDim.x + threadIdx.x;
    for (int i = tid; i < len; i += numThreads) {
        output[i] = alpha * input[i];
    }
}

int main() {
    ...
    // Memory allocation on MPU
    mpu_malloc(mpu_input_vec, len * sizeof(float));
    mpu_malloc(mpu_output_vec, len * sizeof(float));
    // Transfer the input data to MPU
    mpu_memcpy(mpu_input_vec, host_input_vec,
               len * sizeof(float), Host2Device);
    // Launch kernel for the computation on MPU
    ScalarVectorMultiply<<<GridCfg, BlockCfg>>>(
        mpu_input_vec, mpu_output_vec, alpha, len);
    // Transfer MPU computation results
    mpu_memcpy(host_output_vec, mpu_output_vec,
               len * sizeof(float), Device2Host);
    ...
}
\end{lstlisting}

Sec.~\ref{sec:prog_interface} introduces the role of MPU in a heterogeneous platform and its programming interface. 
Sec.~\ref{sec:compiler} details the compiler support of transformations from high-level CUDA kernels to optimized programs running on the MPU. 

\subsection{Programming Interface}~\label{sec:prog_interface}

MPU acts as a standalone accelerator in a heterogeneous platform with a similar usage of the GPU.
In terms of the memory system abstraction, MPU has its own memory space independent from the host.
In order to use MPU as an accelerator, the host is responsible to allocate the device memory on MPU, transfer input data to MPU, launch computation kernel to MPU, and transfer computation results from MPU.
Listing~\ref{lst:example} is a code example of offloading scalar-vector multiplication to MPU.
The main function contains code to handle the memory allocation ($mpu\_malloc$), memory transfers ($mpu\_memcpy$), and the kernel launch.
These functions are implemented in MPU runtime for the communication between the host and MPU.
During the kernel launch, MPU runtime is also responsible to dispatch the workload of thread blocks to MPU cores according to the thread block configurations.
To ease the burden of implementing kernels on MPU, MPU supports CUDA programming language as a realization of the SIMT programming model. 
The $\_\_global\_\_$ function in line 2 of Listing~\ref{lst:example} for implementing the scalar-vector multiplication totally follows the CUDA syntax.
As a result, we can leverage a GPU compiler as our front-end to parse the source code and generate intermediate instructions.
Then, our compiler backend as detailed in Sec.~\ref{sec:compiler} is responsible for backend optimizations tailored for MPU architecture.

\begin{algorithm}[!t]
\begin{algorithmic}
\caption{Location annotation algorithm}
\label{algm:loc_annotation}
\footnotesize
\STATE \textbf{Input}: A kernel with a list of instructions $I$
\STATE \textbf{Output}: A location table $L$ for all registers and instructions. 

\STATE $L(reg)$: the location of an register $reg$.
\STATE $L(instr)$: the location of an instruction $instr$.
\STATE Init the location of all registers to unknown $U$.
\STATE Init the location of all instructions to unknown $U$.
\STATE Init the set of registers $R=\emptyset$.
\STATE // Annotate the initial location to address registers, value registers, and 
\STATE // control registers.
\FOR{$instr$ in $I$}
    \FOR{$reg \in \lbrace instr.SrcRegs \bigcup instr.DstRegs \rbrace$}
        \STATE $R = R \bigcup \lbrace reg \rbrace$
    \ENDFOR
    \IF{$instr.type$ $\in$ $Instr_{jump}$}
        \STATE $L(instr.SrcRegs) = F$
    \ENDIF 
    \IF{$instr.type == ld.global$}
        \STATE $L(instr.SrcRegs) = F$
        \STATE $L(instr.DstRegs) = N$
    \ENDIF
    \IF{$instr.type == st.global$}
        \STATE $L(instr.SrcRegs) = N$
        \STATE $L(instr.DstRegs) = F$
    \ENDIF
    \IF{$instr.type \in \lbrace ld.shared, st.shared \rbrace$}
        \STATE $L(instr.SrcRegs) = N$
        \STATE $L(instr.DstRegs) = N$
    \ENDIF
\ENDFOR
\STATE // Propagate the location of known registers to others.
\WHILE{$\forall reg \in R, L(reg)$ does not change}
    \FOR{$instr$ in $I$}
        \FOR{$reg$ in $instr.SrcRegs$}
            \IF{$L(reg) == U$}
                \STATE $L(reg) = L(instr.DstRegs)$
            \ENDIF
            \IF{$L(reg) != L(instr.DstRegs)$}
                \STATE $L(reg) = B$
            \ENDIF
        \ENDFOR
    \ENDFOR
\ENDWHILE
\STATE // Annotate the location of instructions according to the location of their 
\STATE // destination registers.
\FOR{$instr \in I$}
    \STATE $L(instr) = L(instr.DstRegs)$
\ENDFOR
\end{algorithmic}
\end{algorithm}

\begin{figure}[!t]
  \centering
  \includegraphics[width=1\linewidth]{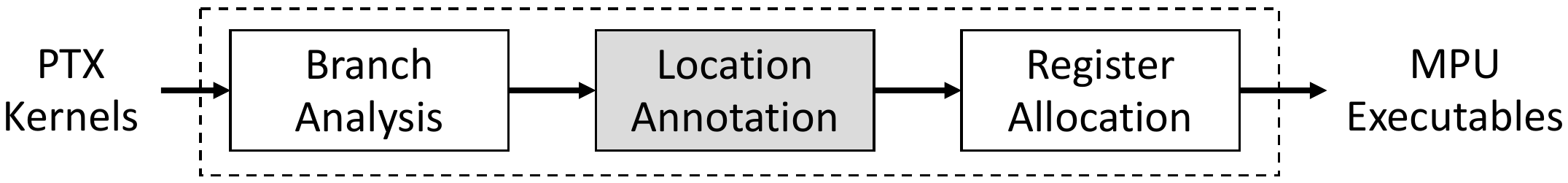}
  \caption{The backend of MPU compiler generating MPU executable kernels from PTX kernels.}
  \label{fig:compiler:backend}
\end{figure}

\subsection{MPU Compiler}~\label{sec:compiler}

To enable the SIMT programming model, MPU supports an end-to-end compilation flow from CUDA programs to MPU executable programs. 
This compilation flow contains a novel static analysis stage to optimize the location assignment of instructions by reducing data movement between MPU cores and near-bank units.
Experimental results in Sec.\ref{subsec:exp:compiler_analysis} demonstrate that this novel location annotation improves the performance of MPU compared with a default hardware policy. 

The end-to-end compilation flow of MPU includes frontend stages and backend stages.
In frontend stages, the MPU compiler reuses $nvcc$~\cite{nvidia2013compiler} to compile CUDA programs~\cite{kirk2007nvidia} to generate kernels in Parallel Thread Extension (PTX)~\cite{PTX} ISA which is a kind of intermediate representation of CUDA kernels.
Then, the backend generates the MPU hardware executables from the PTX kernels, which includes three main stages as shown in Fig.\ref{fig:compiler:backend}.
Among these three stages, the branch analysis and register allocation stages are common to support the SIMT programming model.
The branch analysis stage infers the re-convergence point of each jump instruction so that the hardware can maintain a SIMT stack to handle thread divergence efficiently during the execution~\cite{fung2007dynamic}. 
This problem can be formulated as the post-dominator analysis of a control-flow graph representing the program. 
The register allocation stage analyzes the liveness of each virtual register in the program to build a register interference graph. 
The allocation of physical registers can then be formulated as a graph coloring problem on this register interference graph. 

\begin{figure}[!t]
  \centering
  \includegraphics[width=1\linewidth]{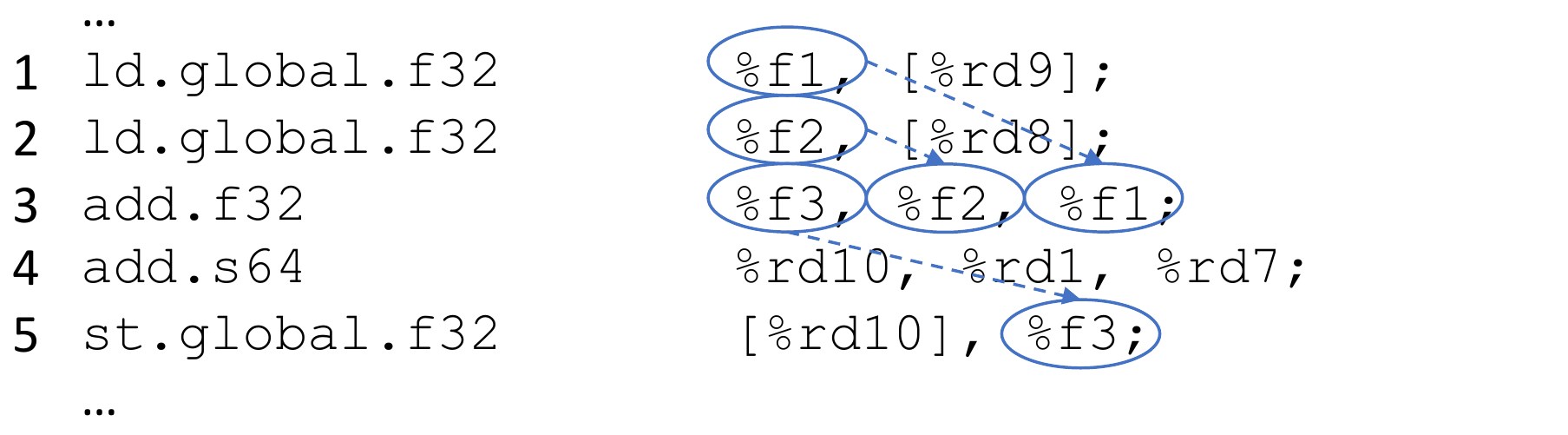}
  \caption{A code snippet of PTX instructions with a dependency chain of near-bank registers highlighted.}
  \label{fig:compiler:code-block}
\end{figure}

The location annotation is a novel backend stage to optimize the performance by annotating the location of instructions as either the near-bank NBU or the far-bank subcore on the base logic die.
As shown in Fig.\ref{fig:instr_offld}, when executing the annotated kernels on MPU, the locations annotated on instructions will be used to finalize the runtime instruction offloading decision as explained in Sec.\ref{subsec:arch:instr_offld}.
The main idea of this optimization is a heuristic approach based on the static analysis extracting the dependency chains of the control-related, address-related, and value-related registers. 
First, value-related registers will be annotated as near-bank registers while control-related and address-related registers will be annotated as far-bank registers.
Then, instructions from the dependency chain of value-related registers are annotated as the near-bank instructions while the rest of instructions are annotated as the far-bank instructions. 
For example, Fig.\ref{fig:compiler:code-block} highlights a dependency chain of value registers.
Offloading the compute instruction in line 3 to near-bank compute logic can eliminate the data transfer on TSVs for registers \%f1, \%f2, and \%f3.
Finally, the annotated program will be passed to the register allocation stage where registers annotated as different locations will not share the same physical register.

The static analysis of the location annotation is implemented as an iterative algorithm shown in Algorithm~\ref{algm:loc_annotation} to realize the idea of decoupling register dependency chains.
Initially, the location of all registers is annotated as $U$ (unknown). 
The value registers of the global load/store instructions are annotated as $N$ (near-bank) while the address registers of these instructions are annotated as $F$ (far-bank) because of the LSU design.
In addition to that, the location of predicative registers is annotated as $F$ (far-bank) because control-related instructions are executed on the far-bank pipeline stages. 
Then, our algorithm iteratively scans over the program to update register locations.
In particular, if the destination register location is known for an instruction, its source registers will follow the same location annotation.
If a register is annotated as both $N$ and $F$ from different instructions, it will be annotated as $B$, meaning that this register could appear on both far-bank and near-bank pipeline stages.
This process will finish once the annotated locations of all registers converge. 
Finally, the location of instruction follows the same location as its destination register. 

\section{Evaluation}\label{sec:exp}
In Sec.\ref{subsec:exp:setup}, we introduce the experiment setup and methodologies.
In Sec.\ref{subsec:exp:ppat}, we show the performance, area, energy, and thermal results of MPU.
In Sec.\ref{subsec:exp:hw_analysis}, we demonstrate the benefit of MPU's architecture optimizations and the comparison with the prior processing-on-base-logic-die designs.
In Sec.\ref{subsec:exp:compiler_analysis}, we present the effectiveness of the location annotation in our compiler backend optimization.

\subsection{Experimental Setup}\label{subsec:exp:setup}
\begin{table}[!t]
    \centering
    \scriptsize
    \caption{The workloads of the benchmark suite.}
    \label{tab:exp:benchmark}
    \begin{tabular}{|l|l|l|l|}
    \hline
        Workload & App Domain & Reference & Description \\
    \hline 
        BLUR & Image Processing & Halide~\cite{ragan2013halide} & 3x3 blur. \\
        CONV & Machine Learning & TensorFlow~\cite{abadi2016tensorflow} & 3x3 conv. \\
        GEMV & Linear Algebra & cuBLAS~\cite{cuBLAS} & Matrix-vector multiply. \\
        HIST & Image Processing & CUB~\cite{CUB} & Histogram. \\
        KMEANS & Machine Learning & Rodinia~\cite{che2009rodinia} & K-means clustering. \\
        KNN & Machine Learning & Rodinia~\cite{che2009rodinia} & K-nearest-neighbour. \\
        TTRANS & Linear Algebra & cuBLAS~\cite{cuBLAS} & Tensor transposition. \\
        MAXP & Machine Learning & TensorFlow~\cite{abadi2016tensorflow} & Max-pooling. \\
        NW & Bioinformatics & Rodinia~\cite{che2009rodinia} & Sequence alignment. \\
        UPSAMP & Image Processing & Halide~\cite{ragan2013halide} & Image upsample. \\
        AXPY & Linear Algebra & cuBLAS~\cite{cuBLAS} & Vector add. \\
        PR & Linear Algebra & CUB~\cite{CUB} & Parallel reduction. \\
    \hline 
    \end{tabular}
\end{table}

\noindent \textbf{Benchmark.}
To evaluate the effectiveness of the MPU design in supporting data-intensive parallel programs, we select a set of representative CUDA workloads as shown in Table.\ref{tab:exp:benchmark}. 
In particular, these workloads are from various important application domains including image processing, machine learning, linear algebra, and bioinformatics. 
Because our MPU compiler needs either CUDA source code or PTX kernels to generate MPU executable programs, we have CUDA implementations of these workloads from either well-known GPU benchmarks, such as Rodinia~\cite{che2009rodinia}, or writing CUDA programs in the same functionality while achieving performance comparable to state-of-the-art libraries, such as cuBLAS~\cite{cuBLAS}. 

\begin{table}[!t]
\scriptsize
\centering
\caption{MPU hardware configuration parameters.}
\label{tab:hw_config}
\begin{tabular}{|l|l|}
\hline
Parameter Names                         & Configuration       \\
\hline
\hline
Proc/(3D,Core)/(Subcore,NBU/Bank/RowBuf) & 8/(4,16)/(4,4/4/4) \\
SIMT/BankIO/TSV/(on)offchip\_bus (Bit) & 32/256b/1024/(256)128 \\
Bank/Icache/(Far)Near-bank RF/Smem (Byte)         &  16M/128K/(32K)16K/64K \\
\hline
tRCD/tCCD/tRTP/tRP/tRAS/tRFC/tREFI~\cite{kim2015ramulator} & 14/2/4/14/33/350/3900 \\
fCore  / fTSV / fRouter / f(on)offchip\_bus (GHz) & 1/2/2/(2)2 \\
\hline
RD,WR/PRE,ACT/REF/RF/SMEM~\cite{chen2012cacti} (J/access) & 0.15n/0.27n/1.13n/40.0p/22.2p \\
Operand\_collector / LSU-Extension (J/access) & 41.49p/39.67p \\
TSV~\cite{o2017fine} / (on)off-chip bus~\cite{chen2012cacti,pugsley2014ndc} (J/bit) & 4.53p/(0.72p)4.50p  \\
\hline
DRAM\_rowbuffer\_policy / DRAM\_schedule & open\_page / FR-FCFS \\
\hline
\end{tabular}
\end{table}

\textbf{Hardware Configuration.}
Using the 3D-stacking memory configuration similar to the previous near-bank accelerators~\cite{shin2018mcdram,yazdanbakhsh2018dram,aga2019co,gu2020ipim}, MPU needs no changes to DRAM's core circuit except the multiple activated row-buffers enhancement~\cite{kim2012case}.
The detailed hardware configuration, latency values, energy consumption, and DRAM settings are presented in Table.\ref{tab:hw_config}.
MPU contains $8$ processors (total $\sim926mm^2$) to compare with a Tesla V100 GPU card~\cite{Ref:GPU} with $4$ HBM  stacks (total $\sim1199mm^2$), where one HBM stack consumes $\sim96mm^2$ footprint~\cite{sohn20171}.

\textbf{Simulation Methodology.}
We develop an event-driven simulator using SimPy~\cite{matloff2008introduction}, which adapts the simulation framework from GPGPU-Sim~\cite{bakhoda2009analyzing} for SIMT core model, Ramulator~\cite{kim2015ramulator} for DRAM model, and Booksim~\cite{jiang2013detailed} for interconnect network model.
MPU is designed to run at a clock frequency of 1GHz under the 20nm technology node.
For \underline{energy and latency modeling}, we first use the design compiler to get the power values for the SIMT core pipeline based on Harmonica project~\cite{kersey2017lightweight}.
Then, we use cacti~\cite{chen2012cacti} to evaluate the register file, shared memory, and the DRAM bank.
Since the major components in the operand collector and the LSU-Extension are SRAM buffers, we also use cacti to evaluate their latency and energy values.
We model the router latency and energy consumption using BookSim2's model~\cite{jiang2013detailed}, and the TSV and on/offchip buses adopt parameters from previous studies~\cite{o2017fine,chen2012cacti,pugsley2014ndc}.
For the ALU, we use the measured results from PTX instructions~\cite{arafa2019low,arafa2020verified}.
For \underline{area evaluation}, we use design compiler~\cite{dupenloup2004automatic} to analyse pre-layout area of the vector ALU and the SIMT core pipeline~\cite{kersey2017lightweight}.
We use AxRAM's area result~\cite{li2015rram} for the in-dram memory controller and scale it to 20nm.
The area for the shared memory, register file, operand collector, and LSU-Extension are derived from cacti~\cite{chen2012cacti}.
For all the above-evaluated components on the DRAM die, we conservatively assume $\times2$ area overhead considering the reduced number of metal layers in the DRAM process~\cite{yazdanbakhsh2018dram}.
For multi-row-buffer support, we include the overhead of $128$ extra row address latches~\cite{kim2012case} per memory controller to enable simultaneous activations of $4$ subarray row buffers.
For the \underline{GPU evaluation}, the GPU performance and power results are collected with the help of \emph{nvprof} and \emph{nvidia-smi}, respectively.

\subsection{Performance, Area, Energy, and Thermal Analysis}\label{subsec:exp:ppat}

\begin{figure}[!t]
  \centering
  \includegraphics[width=1\linewidth]{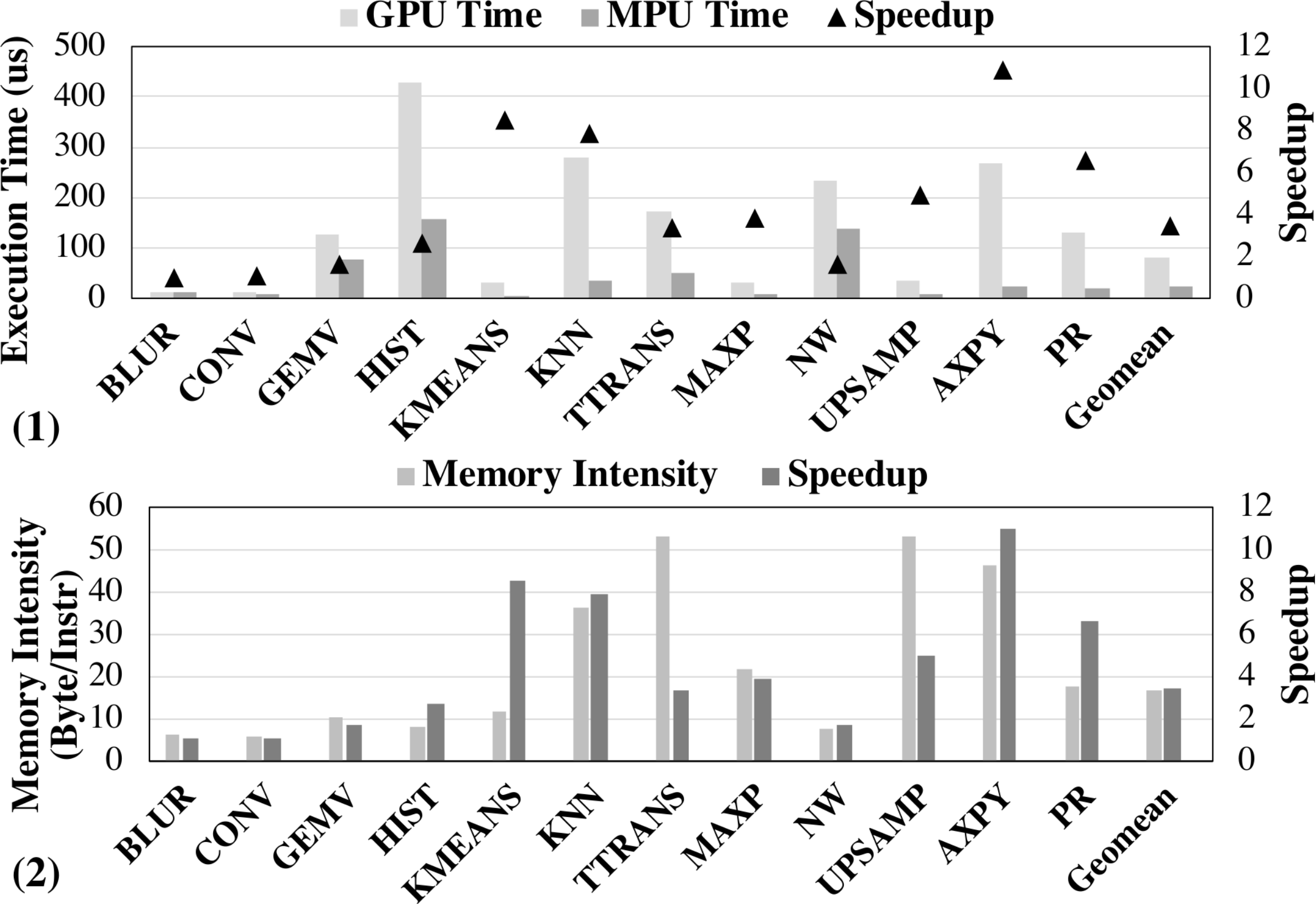}
  \caption{(1) Execution time and speedup comparison with the GPU. (2) Workloads memory intensity and speedup.}
  \label{fig:performance}
\end{figure}

\textbf{Performance.}
MPU achieves $3.45\times$ speedup on average over the GPU as shown in Fig.\ref{fig:performance} (1).
This speedup is contributed by the improved memory bandwidth from the hybrid-pipeline near-bank architecture, the architecture optimizations for the SIMT programming model (Sec.\ref{subsec:exp:hw_analysis}), and the compiler optimizations for the locations of instructions (Sec.\ref{subsec:exp:compiler_analysis}).

To further explain different speedup numbers across workloads, we plot the memory intensity (Byte/Instruction) and the speedup of these workloads in Fig.\ref{fig:performance} (2).
First, we observe that the speedup number has a strong correlation with the memory intensity because the memory intensity represents the demand of workloads for memory bandwidth.
As MPU provides more memory bandwidth than the GPU ($4.13\times$ in measurement), for benchmarks with simple memory access and compute patterns (e.g., AXPY), the speedup is proportional to the memory intensity.
Second, we find that some benchmarks show higher (KMEANS) and lower (TTRANS, UPSAMP) speedup numbers than their memory intensity. 
The reason is that memory dependency cannot reflect memory latency characteristics and complex program behaviors.
For KMEANS, MPU provides additional latency reduction compared to the GPU, as the compute instructions are mostly data-dependency free so the performance is less sensitive to the number of instructions.
For TTRANS and UPSAMP, complicated control flow and data-dependency hinder the memory parallelism, so the abundant memory bandwidth in MPU is not fully utilized.

\begin{table}[!t]
   \centering
   \scriptsize
   \caption{Area evaluation of MPU components on the DRAM die considering DRAM process overheads.}
   \label{table:area}
   \begin{tabular}{|l|l|l|l|}
        \hline
        Name & Number &  Area Per Die ($mm^2$) & Overhead ($\%$)\\
        \hline
        \hline
        Shared Memory & 4 & 0.84 & 0.88 \\
        Register File & 16 & 9.71 & 10.12 \\
        Memory Controller & 16 & 0.63 & 0.66 \\
        Operand Collector & 64 & 2.43 & 2.53 \\
        Vector ALU & 16 & 3.74 & 3.90 \\
        LSU-extension & 16 & 2.43 & 2.53 \\
        Multi-row-buffer Support & 64 & 0.01 & 0.01 \\
        \hline
        Total & - & 19.80 & 20.62 \\
        \hline
   \end{tabular}
\end{table}

\textbf{Area.}
MPU's hybrid pipeline architecture is area-efficient because only a small part of the pipeline backend components are added in the DRAM die, saving the area for other pipeline units.
In Table.\ref{table:area}, we evaluate the area of added components and normalize the total overhead to a DRAM die ($96mm^2$~\cite{sohn20171}).
Thanks for our compiler optimizations (Sec.\ref{subsec:exp:compiler_analysis}) which significantly reduce the near-bank register usage, we shrink the near-bank register file to half the size of the far-bank register file.
This brings the total area overhead from $30.74\%$ to $20.62\%$.
We argue this overhead is small for a general purpose SIMT processor, comparing to $10.71\%$ area overhead in previous work which only supports domain-acceleration~\cite{gu2020ipim}.
According to the synthesis result of Harmonica~\cite{kersey2017lightweight} scaled to $20nm$, the $3.4mm^2$ area of the SIMT core with an instruction cache, operand collectors, and a load-store unit can perfectly fit into the available area ($3.5mm^2$~\cite{gao2017tetris}) on the base logic die.
On the contrary, if the whole core is placed in the DRAM die, the total area overhead will increase significantly ($2\times$ compared with the hybrid pipeline of MPU).

\begin{figure}[!t]
  \centering
  \includegraphics[width=1\linewidth]{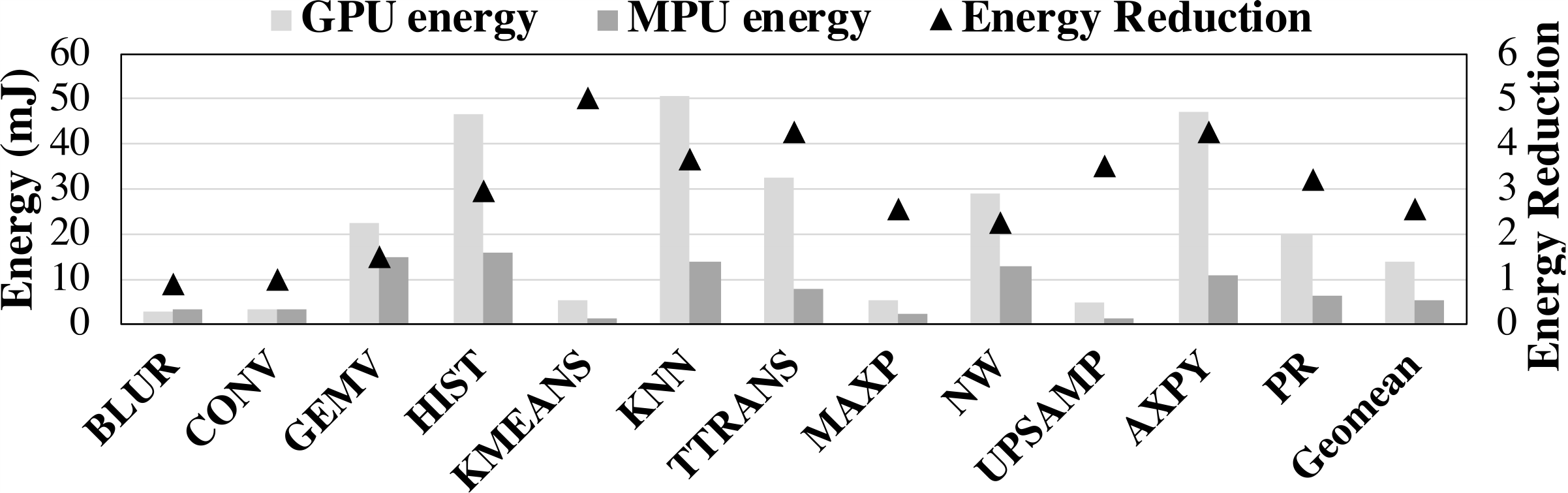}
  \caption{Energy and energy reduction comparison with the GPU.}
  \label{fig:energy}
\end{figure}

\begin{figure}[!t]
  \centering
  \includegraphics[width=1\linewidth]{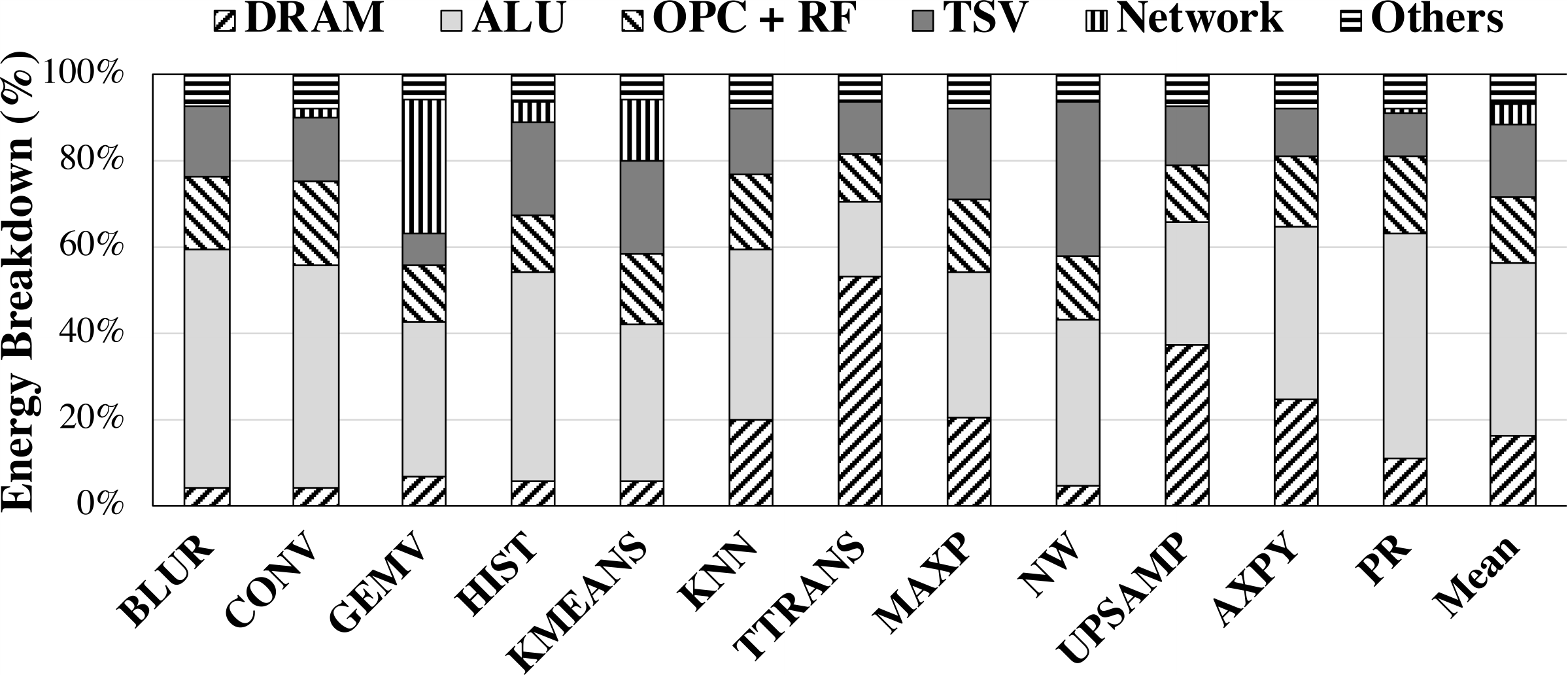}
  \caption{MPU energy breakdown.}
  \label{fig:energy_breakdown}
\end{figure}

\textbf{Energy.}
MPU achieves $2.57\times$ energy reduction on average over the GPU (Fig.\ref{fig:energy}).
The energy reduction mainly comes from the reduction of expensive data movement compared with the GPU, since MPU has a much shorter and simpler data path to access a core's local DRAM banks.
Compared with the complex data path components in the GPU, where the data needs to travel through the TSVs inside the HBM, off-chip links, L2 cache, crossbar network, and then L1 cache to the local register file, the MPU directly offloads the instruction to the DRAM dies to transfer data between the near-bank register file and the DRAM banks.
Also, we observe that for each benchmark the energy reduction in Fig.\ref{fig:energy} is approximately proportional to the speedup in Fig.\ref{fig:performance} (1).
This is because MPU's increased bank-level bandwidth is a result of near-bank data access, which also contributes to the reduction of data movement energy.

In order to further analyze the energy consumption, we provide a detailed energy breakdown in Fig.\ref{fig:energy_breakdown}.
We discover that most of the energy in MPU ($92.94\%$) is spent on computation (ALU consumes $39.82\%$), data access ($31.90\%$), and data movement ($21.22\%$).
The data access energy contains local register file access (operand collectors (OPC) and register file (RF) consume $15.47\%$) and DRAM accesses ($16.42\%$).
For data movement, the energy spent on remote data movement (Network consumes $4.43\%$) is significantly smaller than the local data movement (TSV consumes $16.79\%$).
This well explains the data movement saving advantages of MPU compared to GPU to achieve great energy reduction.

\textbf{Thermal Analysis.}
MPU's peak power is $83W$ per processor considering both DRAM dies and the base logic die, and the peak power density is $552mW/mm^2$. 
The normal operating temperature for HBM2 DRAM dies is $105^{\circ}C$~\cite{sohn20171}, and we conservatively assume the DRAM dies in our case operates under $85^{\circ}C$.
A prior study on 3D PIM thermal analysis~\cite{zhu2016integrated} shows that active cooling solutions can effectively satisfy this thermal constraint ($85^{\circ}C$).
Both commodity-server active cooling solution~\cite{milojevic2012thermal} (peak power density allowed: $706mW/mm^2$) and high-end-server active cooling solution~\cite{eckert2014thermal} (peak power density allowed: $1214mW/mm^2$)) can be used.

\subsection{Architecture Analysis}\label{subsec:exp:hw_analysis}

\begin{figure}[!t]
  \centering
  \includegraphics[width=1\linewidth]{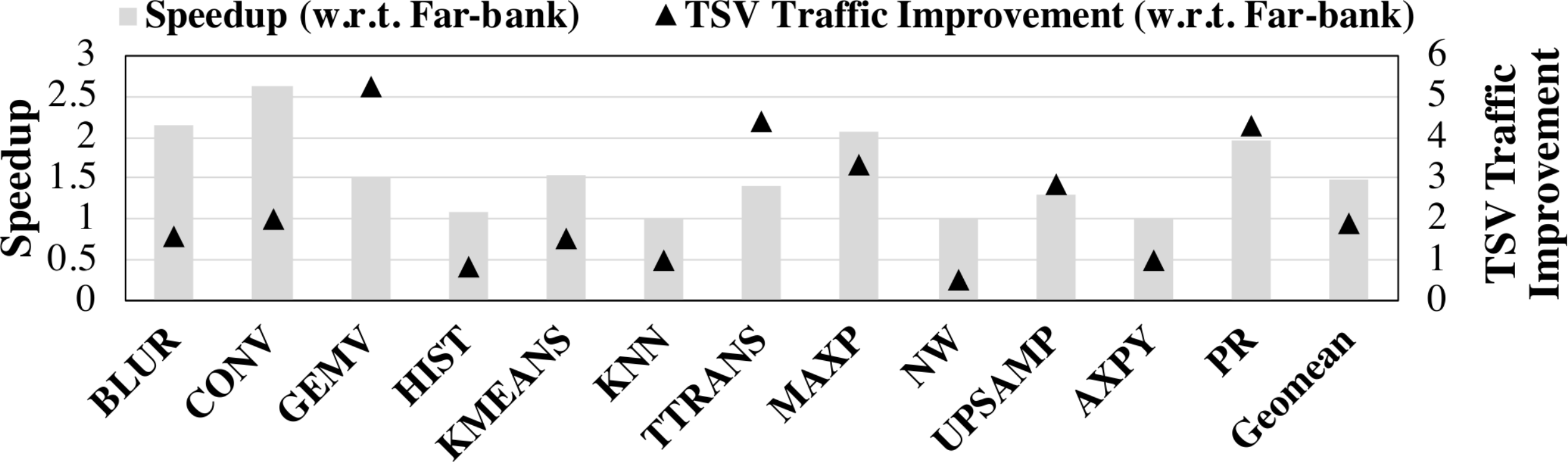}
  \caption{Comparison of near-bank / far-bank smem.}
  \label{fig:exp_smem}
\end{figure}
\textbf{Shared memory optimization.}
To understand the benefit of our near-bank shared memory, Fig.\ref{fig:exp_smem} shows performance results compared with placing the shared memory on the base logic die, denoted as far-bank shared memory. 
In the same figure, we also plot TSV traffic improvement of near-bank shared memory design w.r.t. far-bank shared memory design.
On average, near-bank shared memory design achieves $1.48\times$ speedup and $1.89\times$ TSV traffic improvement compared with far-bank shared memory design.
The performance benefits of near-bank shared memory come from the extensive use of shared memory.
If the shared memory location is far-bank, the contents of near-bank registers need to be brought down to the base logic die for the inter-thread communication through shared memory. 
This will create a lot of register movement traffic and congest the TSVs.
For the near-bank shared memory design, the default locations of value registers for $ld/st.global$ and $ld/st.shared$ are all near-bank.
Thus less register movement will be involved, easing the bandwidth pressure on the TSVs.
However, since the number of instructions offloaded to NBUs also rises, this may increase TSV traffic, as we observe that for some workloads with speedup larger than 1, the TSV traffic improvement may be slightly less than 1 (HIST, NW).
For workloads that do not use shared memory, both the performance and TSV traffic are identical to the location of shared memory.

\begin{figure}[!t]
  \centering
  \includegraphics[width=1\linewidth]{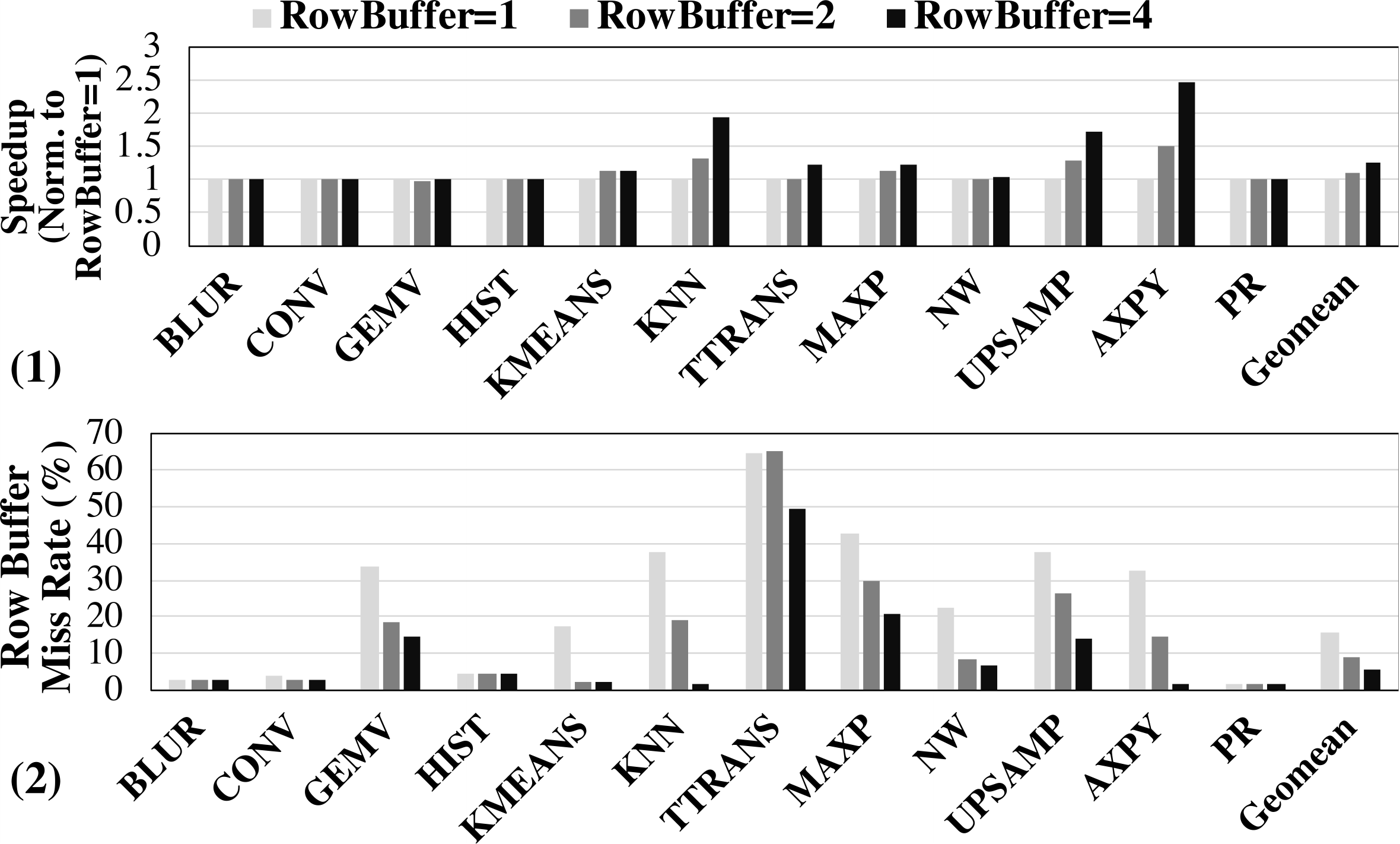}
  \caption{Comparison of the number of activated row-buffers on (1) performance and (2) row-buffer miss rate}
  \label{fig:exp_rb}
\end{figure}

\textbf{Multiple activated row-buffers analysis.}
To understand the benefits of multiple activated row-buffers, we compare the performance of all workloads running on MPU with different numbers of activated row-buffers.
Fig.\ref{fig:exp_rb} shows such performance comparisons where the speedup is normalized to a single row-buffer.
As shown in the Fig.\ref{fig:exp_rb} (1), the speedup numbers are $1.10\times$ and $1.25\times$ when we increase the number of activated row-buffers to $2$ and $4$, respectively.
The row-buffer miss rate in Fig.\ref{fig:exp_rb} (2) indicates that as we increase activated row-buffer numbers to $2$ and $4$, the miss rate reduces from $15.60\%$ to $9.20\%$ and $5.45\%$, respectively.
Because more activated row-buffers can effectively reduce the row-buffer ping-ping effect in the dynamic scheduling of warps, increasing the number of activated row-buffers effectively reduces average DRAM access latency to improve end-to-end time.
Especially, we observe that KNN, UPSAMP, and AXPY significantly benefit from the increased number of activated row-buffers due to severe ping-pong effects on a single row-buffer.

\begin{figure}[!t]
  \centering
  \includegraphics[width=1\linewidth]{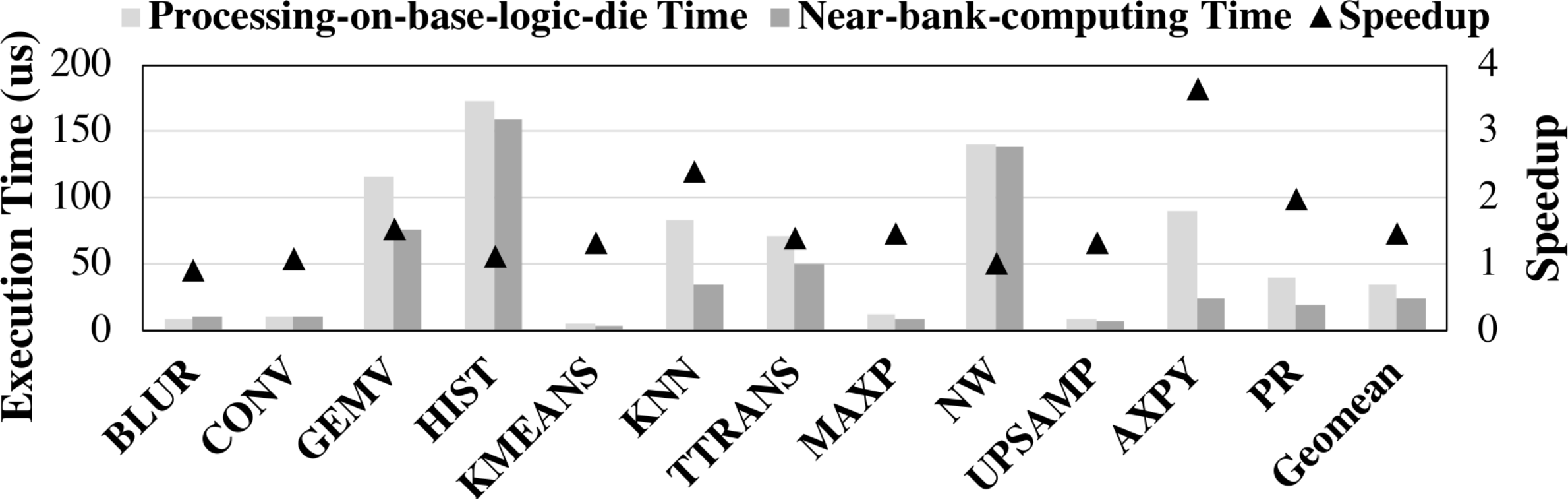}
  \caption{Execution time and speedup comparison with the processing-on-base-logic-die solution.}
  \label{fig:mpu_pob}
\end{figure}
\textbf{Comparison with processing-on-base-logic-die (PonB) solution.}
We compare MPU with the state-of-the-art general purpose near-data SIMT processors by placing all compute logic on the base logic die, denoted as PonB.
The end-to-end execution time shown in Fig.\ref{fig:mpu_pob} demonstrates that on average MPU achieves $1.46\times$ speedup up compared with the PonB solution.
This performance improvement is contributed by a significant amount of instructions offloaded for near-bank computations.
This reduces data movements on the TSVs which have a much lower bandwidth than bank-level memory bandwidth.

\subsection{Effectiveness of Compiler Optimizations}\label{subsec:exp:compiler_analysis}

\begin{figure}[!t]
  \centering
  \includegraphics[width=1\linewidth]{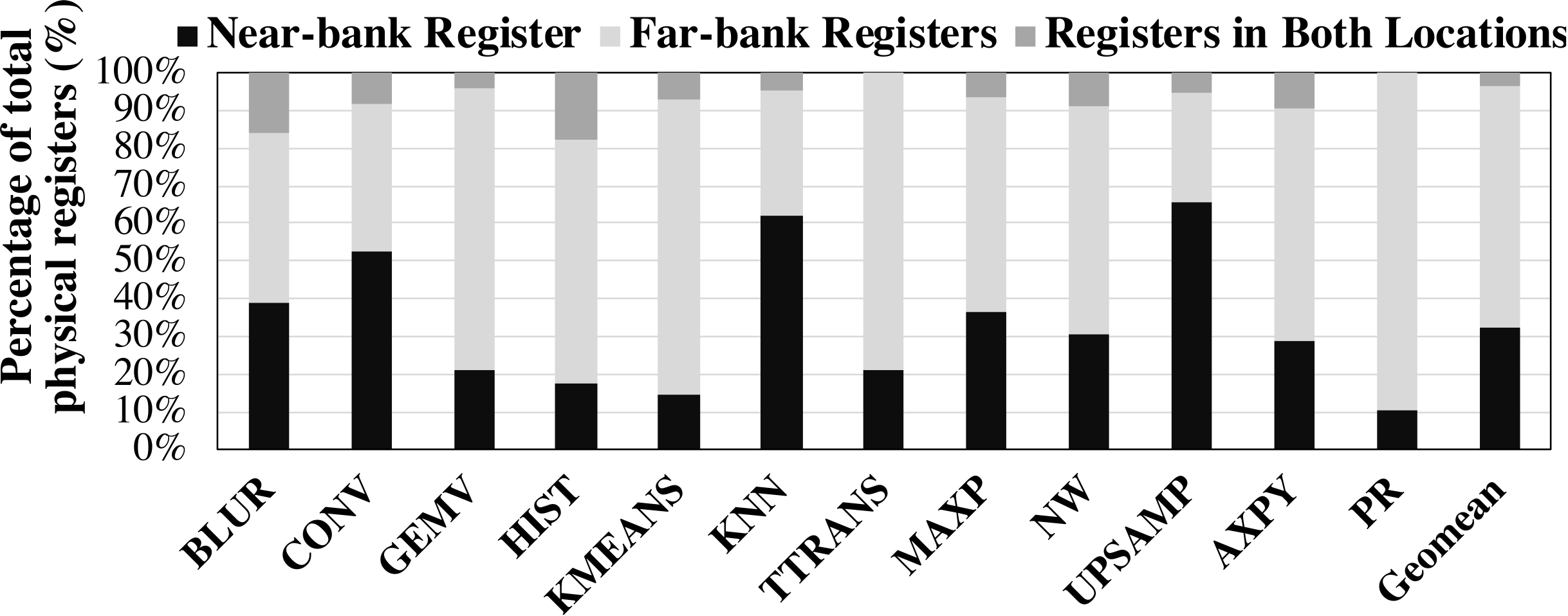}
  \caption{Register location analysis.}
  \label{fig:register_breakdown}
\end{figure}

\begin{figure}[!t]
  \centering
  \includegraphics[width=1\linewidth]{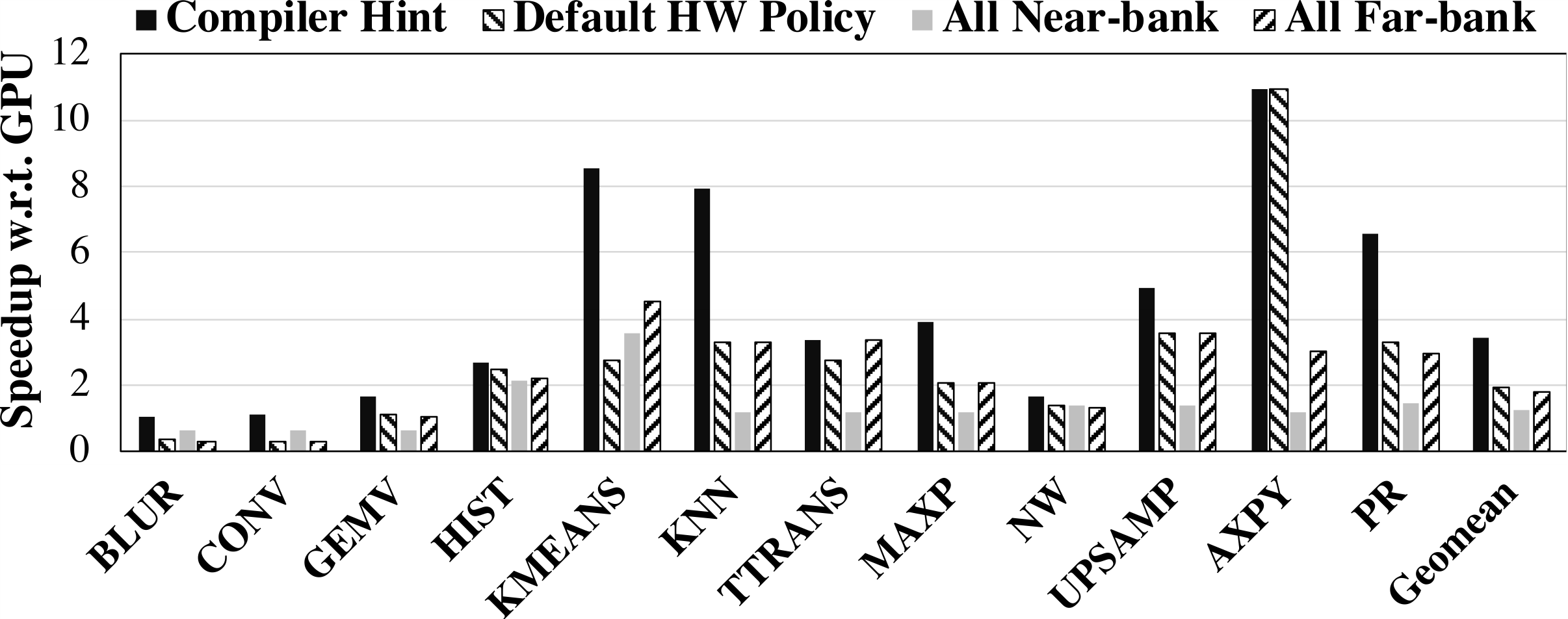}
  \caption{Comparison of performance for different instruction location policies.}
  \label{fig:inst_location_perf}
\end{figure}

We first conduct static analysis according to the iterative algorithm introduced in Sec.\ref{sec:compiler} to infer the locations of registers.
The breakdown of registers on different locations for all workloads is shown in Fig.\ref{fig:register_breakdown}. 
On average, $32.5\%$ registers only appear in near-bank locations, $63.7\%$ registers only appear in far-bank locations, and $3.8\%$ registers could appear in both locations.
Because only registers appearing in near-bank locations need to use the near-bank register file, we effectively shrink the size of the near-bank register file to reduce its area overhead.
This breakdown also demonstrates a clear separation of near-bank registers from far-bank registers.
Only a small portion of registers could appear in both locations.
This clear separation comes from a clear separation of two classes of dependency chains.
The first class involves computations on the data value loaded from the DRAM, and the second class involves integer calculations for DRAM addresses and control-flow related variables, such as loop variables.
Usually registers for these two classes of dependency chains do not interfere with each other, as registers associated with the first class usually exist in near-bank locations, and registers related to the second class reside on the base logic die.
Therefore, for most registers (more than $95\%$) we can assign them to a certain near/far-bank location to reduce the register file usage.

We further evaluate the performance of different instruction location policies with the GPU as shown in  Fig.\ref{fig:inst_location_perf}.
On average, using the proposed instruction annotation optimizations, we achieve $3.45\times$ speedup w.r.t. GPU. 
However, using hardware's default instruction location policy, offloading all instructions to near-bank compute-logic, or offloading all instructions to far-bank compute-logic, we achieve $1.92\times$, $1.22\times$, and $1.78\times$ speedup, respectively.
Compared with the default hardware policy and both naive offloading strategies, our instruction location annotation is based on the annotated register location. 
Because of the clear separation of two classes of aforementioned dependency chains, our instruction location annotation assigns most of the computation on data values to near-bank and computation on addresses or control-flow conditions to far-bank.
As a result, the register movement traffic on TSVs is minimized, which eventually boosts the performance of programs running on MPU.

\section{Related Work}\label{sec:related}
\textbf{General Purpose Near-data-processing Platforms:}
Pioneering studies~\cite{kogge1994execube,kang1999flexram,draper2002architecture} attempted to integrate the entire processor on the DRAM die, which incurs considerable area overheads.
Compared with them, MPU only places lightweight components in the DRAM dies through a hybrid pipeline, which significantly reduces the overhead.
Recently, there are a number of practical general purpose near-data-processing solutions that explore near-cache~\cite{lockerman2020livia}, near-memory-controller~\cite{onur2015reducing}, near-DIMM~\cite{asghari2016chameleon,alian2018application}, and 3D-stacking processing-on-logic-die CPU-style~\cite{ahn2015pim,boroumand2016lazypim,cho2020chonda} and GPU-style~\cite{hsieh2016transparent,pattnaik2016scheduling,kim2017toward,wen2017optimizing,wu2020tuning} platforms.
However, these solutions have several drawbacks.
First, they have moderate bandwidth improvement, due to the hierarchically shared bus of the main memory.
To overcome this drawback, MPU unleashes bank-internal bandwidth through near-bank computing.
Second, the communication between the host and the near-data logic introduces extra data traffic because of shared memory space, which may offset the benefit of near-data-processing, including fine-grained instruction offloading overhead~\cite{ahn2015pim,hsieh2016transparent}, cache coherence traffic~\cite{boroumand2016lazypim}, concurrent host access stall~\cite{cho2020chonda}, and inconsistent data layout requirement~\cite{wu2020tuning}.
Different from these prior studies, MPU has an independent memory space and supports end-to-end kernel execution, the same as discrete GPU cards~\cite{macri2015amd,o2014highlights}.

\textbf{Domain-specific Near-data-processing Accelerators:}
A large number of previous studies have explored domain-specific accelerators using near-data-processing ideas, including approximate computing~\cite{yazdanbakhsh2018dram}, image and video processing~\cite{xie2017processing,gu2020ipim}, deep learning~\cite{akin2019case,shin2018mcdram}, graph analytics~\cite{nai2017graphpim}, bioinformatics~\cite{huangfu2019medal}, garbage collection~\cite{jang2019charon}, address translation~\cite{picorel2017near}
and data transformation~\cite{akin2015hamlet}.
These designs usually adopt domain-specific processing logic, customized data paths, and application-tailored software mapping strategies.
The lack of programmability for these accelerators confines them to a niche application market, adding non-recurring engineering costs for silicon manufacturing.
In contrast, the SIMT programming model supported by MPU can benefit a wide range of data-intensive parallel programs, and our end-to-end compilation flow greatly eases the burden of programmers.

\textbf{Analog Process-in-memory Architecture:}
In addition to the digital near-data-processing solutions, recently there is a surge in researches about analog process-in-memory (PIM) architecture~\cite{strukov2008missing,mittal2019survey}.
Different from the digital solution where the memory array and the computing logic are separate, analog solutions modify the memory array to integrate computing functionalities within memory arrays, thus achieving extremely high computation throughput and energy efficiency.
However, these designs suffer from analog noise~\cite{li2015rram}, limited write-endurance issues of non-volatile devices~\cite{imani2019floatpim}, and high overhead of analog-digital converters~\cite{xia2016switched}.
Although analog PIM solutions are promising for certain application domains such as neural network~\cite{shafiee2016isaac,chi2016prime,song2017pipelayer}, they are still challenging for general purpose computing usage.
In comparison, MPU adopts commercially available 3D stacking technologies~\cite{leidel2016hmc,sohn20171} without modifying the DRAM bank's circuit, and this work has demonstrated promising results of MPU on general purpose data-intensive workloads.

\section{Conclusion}\label{sec:conclusion}
This work proposes MPU (\underline{M}emory-centric \underline{P}rocessing \underline{U}nit), the first SIMT processor based on 3D-stacking near-bank computing architecture.
First, we develop a hybrid pipeline where only a small number of hardware components are added on the DRAM dies and the instructions can be offloaded for near-bank computing.
Second, we explore two architectural optimizations for the SIMT programming model, introducing a near-bank shared memory design to reduce data movements, and multiple activated row-buffers designs to increase bandwidth utilization. 
Third, we present an end-to-end compilation flow for MPU based on CUDA with a backend optimization to annotate the location of instructions as either near-bank or base logic die through the static analysis of programs.
The end-to-end evaluation results of MPU on a set of representative benchmarks demonstrate $3.46\times$ speedup and $2.57\times$ energy reduction compared with an NVIDIA Tesla V100 GPU.
We further conduct studies to show the performance improvement of MPU over prior 3D-stacking processors and identify the benefits of MPU's software and hardware optimizations.



\bibliographystyle{IEEEtranS}
\bibliography{isca21-paper/ref/pim_related.bib,isca21-paper/ref/memory.bib,isca21-paper/ref/gpu.bib,isca21-paper/ref/nn.bib,isca21-paper/ref/others.bib}

\end{document}